# Dependence of Energy Thresholds on Laser Radiation Wavelength in Initiation of Heavy Metal Azides

G. Damamme **, V.M. Lisitsyn *, D. Malis **, V.P. Tsipilev*
* National Research University of Resource-Effective
  Technologies «TPU», Tomsk 634050, Russia
** Commissariat a l'energie atomique, Paris 75015, France

*The behavior of silver and lead azides (scaly and threadlike crystals, and compacted powders) in initiation of explosive decomposition by laser pulsed radiation has been investigated. Initiation energy thresholds in irradiation at wavelengths of 1064 nm, 532 nm, 354.7 nm, 266 nm corresponding to four laser radiation harmonics have been measured. Samples both uncovered and covered with a transparent dielectric (a quartz plate) with the compressive force of up to 0.5GPa were initiated. In the transparent spectral region (the first and second laser harmonics) of the heavy metal azide matrix the effect of covering the sample surface with a plate on initiation energy threshold was insignificant, whereas, in the region of bandgap absorption (the fourth harmonic) strong dependence of the initiation threshold on whether the surface under irradiation is uncovered or covered.*

*The results obtained have been considered with the account of the influence of the transparent plate on gas-dynamic unloading of the thermal site, which is formed in the bandgap absorption of laser radiation in a thin near-surface layer.*

**Introduction**

The influence of laser radiation (LR) wavelength on initiation energy thresholds (IET) is supposed to be determined by the following characteristics of the medium: the absorption index $\mu(\lambda_0)$; scattering index $\beta(\lambda_0)$; refraction index $n_0(\lambda_0)$. If regulate the wavelength these characteristics change in wide ranges and different various modes of illumination of the explosive volume can be created. In this way in transition from UV-region to IR-region of the radiation spectrum $\mu$ value for explosive substances (ES) can vary from $10^5$ to $10^{-5}$ cm$^{-1}$, and laser beam penetration depth in the semi-infinite layer of the substance $(1/\mu)$ can vary over a wide range, that can lead to different effects: heating of the surface layer, macro- or microregions of the volume, formation of electron-hole pairs or optical breakdown, etc.

In the transparent region of heavy metals azides (HMAs) IETs were studied in detail [1, 2, 3], the behavior of the compacted lead azide powder as a model HMA sample in irradiation by neodymium laser pulse ($\lambda_0 = 1060$ nm) in particular. Later [4] IETs for different HMAs, scaly and threadlike crystals among them, were determined at this wavelength. The energy threshold density $H_{05}$ for compacted powders was shown to make values of the order of several $mJ/cm^2$, however, for scaly and threadlike crystals it made tens of $mJ/cm^2$. Thus, in the transparent region of ES matrix low-threshold initiation can be observed.

In [5] the behavior of the compacted lead azide powder was studied in irradiation by nitrogen laser pulse ($\lambda_0 = 337$ nm). In initiation with the uncovered surface IET made several $J/cm^2$, when the surface was covered by a quartz plate, it decreased, i.e. in transition from the transparent region to the region of bandgap absorption IET several orders of the value increased. At first sight this is sufficiently paradoxical.

Similar regularity was found in [6, 7] in initiation by eximer laser pulse ($\lambda_0 = 308$ nm, $\tau = 20$ ns). With the uncovered surface of the pellet *0.5 mm* thick IET made *5.0 J/cm²*, and when the surface was covered with the quartz plate densely pressed to the sample IET decreased down to *0.2 J/cm²*. The results of the research made it possible to conclude that the transparent plate strongly influences gas-dynamic unloading of the heating site which in the region of bandgap absorption is located in the near-surface layer of ES. The inhibitory effect of the surface is described in [8] in initiation of $AgN_3$ macrocrystals by eximer laser radiation.

If follow the above-stated at pressures of plate compressing to the sample commensurable with the pressure in the decomposition site IET is expected to tend to its minimal value, at thus, the sensitivity of ES to radiation will be maximal. Determining IET under optimal conditions for explosive decomposition development can produce some additional significant information on the mechanisms of laser pulse initiation of different ES, including HMAs. The research of the behavior of scaly and threadlike HMA crystals in the transparent region and in the region of bandgap absorption, but in comparable conditions of the experiment, is of particular interest. Actions with the harmonics of neodymium laser radiation meet these conditions.

This paper considers determining IET for HMAs with the uncovered surface and surface covered with a transparent plate (at the compressive pressure ~0.5 GPa) in irradiation by four harmonics of YAG:$Nd^{3+}$-laser.

**Experimental Technique and Results**

A laser bench which is an improved variant of the set-up described in [4] was used in the research. The complex provides an opportunity of multiparameter measurements of the processes accompanying explosive decomposition. A general function scheme of measurements is shown in Fig. 1.

Pulsed YAG:$Nd^{3+}$-laser *1* operating at five harmonics was used as the radiation source. The parameters of laser radiation were as follows:
- Pulse duration:
    $\tau_p = 12$ ns ($\lambda_0 = 1064$ nm);
    $\tau_p = 8$ ns ($\lambda_0 = 266$ nm).
- Pulse energy $W_0$:
    *1500 mJ ($\lambda_0 = 1064$ nm)*;
    *700 mJ ($\lambda_0 = 532$ nm)*;
    *250 mJ ($\lambda_0 = 354.7$ nm)*;
    *170 mJ ($\lambda_0 = 266$ nm)*;

*80 mJ ($\lambda_0 = 213$ nm).*
- Operating mode: from a single pulse up to *5 Hz*.
- Structure of the beam: multimode structure with a homogeneous central part.

Laser radiation *1* was split by a quartz Dove prism *2* (reflection coefficient for S-polarization $\rho_s = 0.1$) or interference mirror *2'* ($\rho_s = 0.5$) and directed by interference mirror *3* ($\rho_s = 0.99$) onto the forming stop *4* which cut out the central part of the beam. The quartz objective *5* (with no spherical aberrations) made a demagnified image of the stop on HMA sample surface *6* (the magnification of the projective scheme depends on the distance between the diaphragm *4* and objective *5* and varied from *1.0* to *0.1*). This type of the scheme allowed to form the size of the homogeneous beam from *1 mm* up to *10 µm* at $\lambda_0 = 1064$ nm and up to *3 µm* at $\lambda_0 = 266$ nm with a relative aperture of the objective equal to *0.25* on the sample surface. The obtained focusing was close to the diffraction limit. At this, a high contrast of light exposure on the image edges was provided.

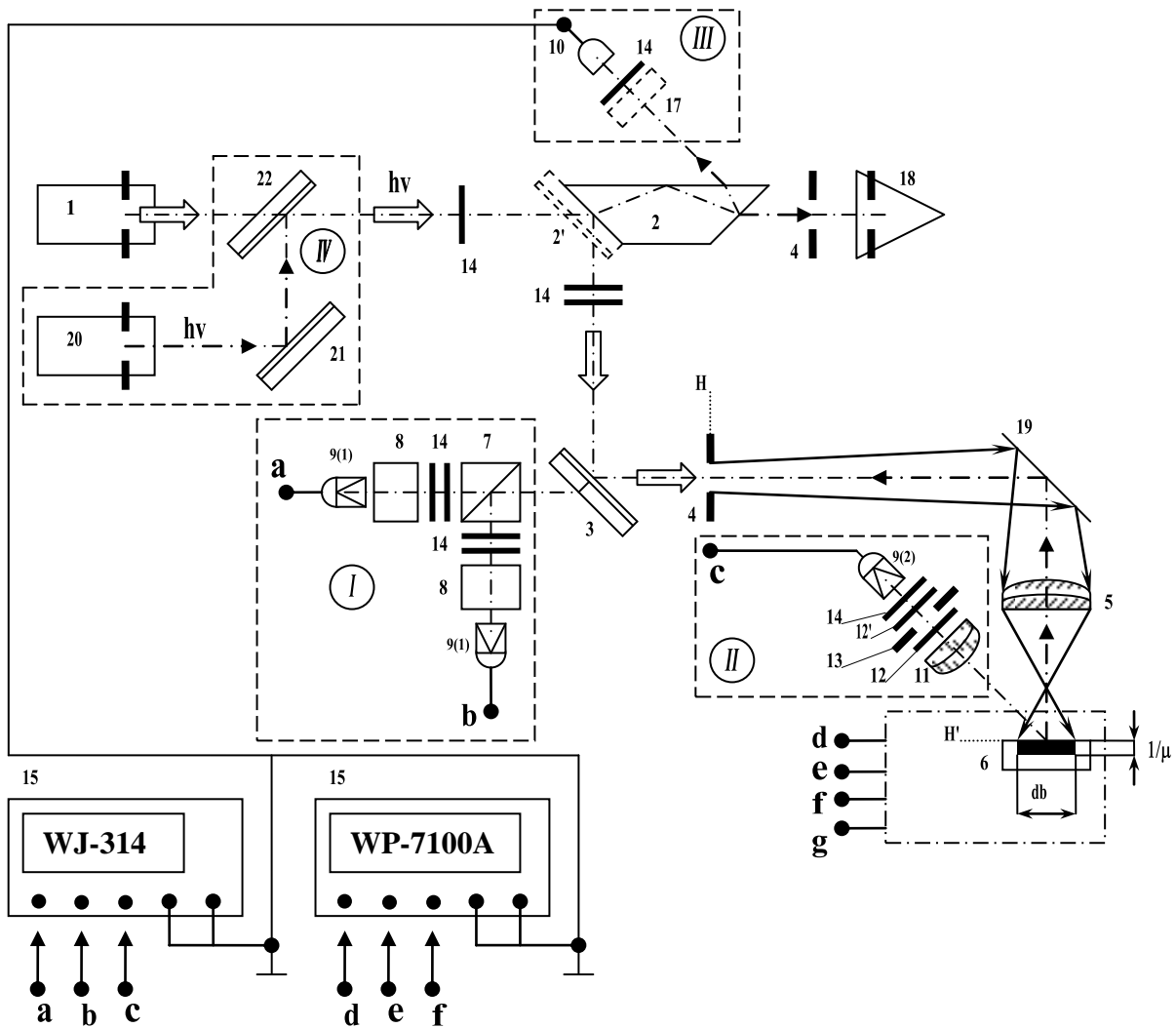

***Fig. 1.*** *Functional scheme of the experimental complex*

*1 is YAG-laser LQ-929 by «SOLAR LS»; 2 is Dove quartz prism; 2' is a replaceable interference mirror for different radiation harmonics with the reflection of 50%; 3 is a replaceable interference mirror with the reflection of 99.9%; 4 is a replaceable metal stop forming the size of the beam on the sample; 5 is a quartz projective objective; 6 is HMA sample (**db** is the size of the beam, $1/\mu$ is the depth of light penetration [cm], $\mu$ is an absorption index [$cm^{-1}$]); 7 is a beam-splitting glass cube; 8 is a small-sized compact monochromator; 9(1, 2) is a photodetector ("Hamamatsu" H5773-04 photomodule); 10 is CP-19 coaxial photodiod; 11 is a projective quartz objective; 12 is a colored light filter; 12' is UVF-2 colored filter; 13 is a field-view stop; 14 is neutral filters; 15 is "LeCroy" WJ-314 (100 MHz) and WP-7100A (1 GHz) digital oscillograph; 16 is measuring replaceable cells; 17 is the phosphor (in operation at the third, fourth and fifth harmonics); 18 is a calorimetric energy meter with an open input; 19 is a metal rotary mirror; 20 is a semiconductor laser; 21 is a rotary mirror; 22 is a dichroic (replaceable) mirror for overlapping the laser beam with a "pilot" beam. **a**, **b**, **c**, **d**, **e** are photodetector outputs, **f** is an output of the acoustic detector. **I** is a registration band channel, **II** is the channel for panoramic observation, **III** is the channel for oscillograph scanning triggering, **IV** is the channel for overlapping the laser and "pilot" beams; H and H' are object and image planes, respectively*

Another advantage of this projective scheme was that it made possible to observe glow from the front surface of the sample and from the region of laser beam action only. This glow was received by the monochromator ***8*** and photodetector ***9(1)*** through the objective *5* and stop *4* which in this case made a spatial filter, then through the mirror *3* and beam-splitting cube *7*. This made it possible to observe glow from the zone of laser action at two wavelengths simultaneously. Measuring the glow intensity by **channel I** (Fig. 1) was made in the region of *400-800 nm* to research the kinetics of explosive decomposition glow.

The observation **channel II** (Fig. 1) made a projective scheme consisting of the quartz objective *11*, field stop *13* and photodetector *9(2)*. The colored filter as filter *12* or interference mirror reflecting one of the harmonics blocked LR reflected from the sample. The colored filter cut the spectral region of *300-400 nm* from the glow spectrum. The scheme made it possible to observe glow from the zone of laser action and beyond its limits. Magnification of the optical scheme was changed within the limits of *5.0-0.5*. This allowed to make the panoramic observation of the region from 1 to *10 mm*. The panoramic channel was adjusted to the maximal sensitivity and was mainly intended to record the thresholds of luminescence, optical breakdown and other preexplosive phenomena, as well as to observe the glow of explosive decomposition products when they are expanding.

The oscillographs were triggered via the **channel III** (Fig. 1). Part of the laser beam was reflected from the Dove prism output face and was received by the vacuum photodiode *10* with the time resolution *~0.1 ns*. The laser pulse was displayed on one of the oscillogramm channels and was used as a reference mark.

Light beams of all the channels were attenuated by neutral filters *14*. To equalize time delays of channels signals were received by photodetectors through optical fibers, and by oscillographs they were received through electric cables of different lengths. Time resolution of photodetectors made *~1 ns*, that of the acoustic detector was equal to *~5 ns*. Time synchronization of channels of all the detectors including detectors of measuring cells *16* was no more than *10 ns*.

The design of the cell (Fig. 2) made it possible to perform measuring under two conditions of laser action. In the first case the irradiated surface of the sample was not covered, in the second one it was covered with a transparent dielectric compressed to it with the force sufficient for retention of decomposition products at the initial stage. This made it possible to exclude gas-dynamic unloading from the sample surface that is essential in explosion initiation in the region of bandgap absorption where absorption index $\mu > 10^4$ $cm^{-1}$. The sample *6* was placed on the glass substrate *23*, *5 mm* thick, covered with a thin mylar film *24*, passing luminous flux only after its decomposition during the sample explosion. The sample with a substrate was placed on the output window (a metal electrode *3 mm* thick) of the acoustic detector *25* to make the acoustic contact (a layer of light oil). The assembly was covered with the quartz plate *26* *16 mm* thick fixed on the top support *27* (made of plexiglass) of the hydraulic press. Radiation from the front surface of the sample was detected by channels *I* and *II* (Fig. 2).

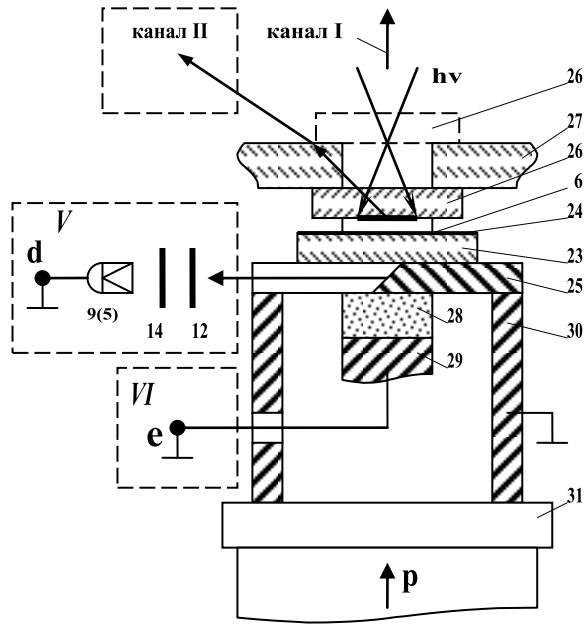

***Fig. 2.*** *The scheme of the cell for measuring IET and kinetic characteristics of explosive decomposition*

*6 is the sample; 9(5) is photodetector ("Hamamatsu" H5773-04 photomodule); 23 is a glass substrate; 24 is a mylar film; 25 is the output window of the acoustic detector with a mirror cut; 26 is a quartz plate; 27 is a press support; 28 is a piezoelement; 29 is an electrode-damper (brass); 30 is the case of the detector; 31 is the hydraulic press piston.* ***II*** *is the channel of panoramic observation;* ***V*** *is the channel of observation from the back side of the sample;* ***VI*** *is the channel for acoustic response detecting*

When the decomposition wave reached the back side of the sample and the film ***24*** destructed (evaporated) there occurred the glow pulse which was received by the recording channel **V** after reflection from the mirror cut in the electrode ***25***. In conditions of surface initiation (initiation of compacted powders or crystals in the region of bandgap absorption) this cell allowed to obtain information on the velocity of decomposition wave propagation deep in the sample.

The signal of the acoustic detector from an electrode-damper ***29*** was received by the recording channel **VI**. The channels were adjusted and synchronized as follows. The sample ***6*** was removed from the cell, then the film ***24*** was pressed to the plate ***26***. The film was irradiated at LR density of $\sim$5-10 J/cm$^2$, that led to its instantaneous (within the time of LR action) explosive evaporation, occurrence of an intensive wideband glow and strong shock compression. Thus, the film explosion was an ideal way to adjust the scheme. The acoustic delay of the detector was found by this technique and it made *1680 ns*.

The research was performed as follows. Threadlike *(0.4×0.1 mm$^2$)* or scaly *(1.5×1.5×0.1 mm$^3$)* silver azide crystals were placed on the glass substrate ***23*** and pressed with the plate ***26*** with the force of *0.5 GPa*. In initiation with the uncovered surface of HMA sample the plate ***26*** was placed to the top plane of the support ***27*** (its position is plotted in a dotted line (Fig. 2)), a ring with the height equal to thickness of the plate and an additional ring with the height equal to the thickness of the sample were put on its place. At this in contraction of the construction the sample was brought in its initial position, i.e. in the plane (*H'*) conjugated with the stop *4* (*H*), and the plate *26* protected the objective from the damage by explosion products. This technique allowed to maintain identical conditions for irradiation of both covered and uncovered samples. The compacted silver and lead azide powders were made in the form of pellets *3 mm* in diameter and *0.2 mm* thick (the compaction pressure made *0.8 GPa*, density was equal to *4.0 g/cm$^3$*). In some cases a mylar or aluminium film (*10 μm* thick) was placed between the pellet and the substrate, that allowed to detect the run of the explosive decomposition wave to its back surface.

In switching from one harmonic to a different one the mirror *3* was replaced, and in operating at the third and fourth harmonics there were placed mirrors *2'*. At each of the harmonics the beam splitting coefficient was measured subject to losses in all the elements of the projective scheme. In all the experiments the diameter of the laser beam on samples surfaces was maintained equal to *1 mm*. To plot the probabilistic curve of initiation (frequency ratio curve of initiation) no less than *25-30* tests were performed in each series of the experiments. According to the technique stated in [1, 9] the energy value *$W_{05}$* or energy density *$H_{05}$* corresponding to *W* or *H* values at which the probability of explosion is equal to *50%* were taken as IET, and the region of probabilistic explosion *ΔH* bounded by the tangent to the curve in the point *$H_{05}$* was accepted as the probabilistic explosion interval. The results of IET measuring are shown in Table 1. Covering of the sample surface in the transparent region (the first and the second laser harmonics) is seen to weakly influence IET, whereas, for the third and fourth laser harmonics IETs depend on whether the sample is covered or uncovered and can differ two orders and more.

*Table 1. Experimental and calculated (by thermal model of initiation) values $H_{05}$ with different laser radiation wavelengths*

| Material | Energy threshold density $H_{05}$, mJ/cm$^2$ (experiment) | | | | | | | |
|---|---|---|---|---|---|---|---|---|
| | $\lambda_0 = 1064$ nm | | $\lambda_0 = 532$ nm | | $\lambda_0 = 354.7$ nm | | $\lambda_0 = 266$ nm | |
| | Uncov. | Cov. | Uncov. | Cov. | Uncov. | Cov. | Uncov. | Cov. |
| SC AgN$_3$ | 52±4.0 | 20±3.0 | 16±5.0 | 9.5±3.0 | 40±20 | 15±5.0 | 1600±350 | 10±5 |
| ThC AgN$_3$ | 24±5.0 | | 6±2.0 | 4.5±1.5 | 600±150.0 | 30±3.0 | 1000±300 | 20±8 |
| CP AgN$_3$ | 8±1.4 | 6±1 | 4±1.1 | 3.5±1.0 | 16±1.8 | 11±1.5 | 350±100 | 10±2 |
| CP Pb(N$_3$)$_2$ | 12±1.2 | 11±1.2 | 7±1.0 | 4.0±0.8 | 10±0.8 | 6.5±1.0 | 700±200 | 10±2 |

*SC is scaly crystal; ThC is threadlike crystal; CP is compacted powder.*

The following features in behavior of HMAs were found in the experiments. At the first harmonic of radiation ($\lambda_0 = 1064$ nm) when the sample explosion failed their optical characteristics did not change. In initiation both within the time of laser action, and within the induction period no sample glow was observed except for the glow of explosive decomposition. The induction period at the initiation threshold made at the average of *100-200 ns* for different types of samples. This type of behavior agrees with the results obtained earlier [4] in radiation in a single-mode regime of generation (*$\lambda_0 = 1064$ nm*).

At the second laser harmonic ($\lambda_0 = 532$ nm) when the sample explosion failed the glow signal with the threshold less than *1 mJ/cm²* could be observed. The time position of the luminescence pulse, its duration and form coincided with LR pulse. At levels of action close to IET (*~10 mJ/cm²*) the luminescence signal was sufficiently intensive and made it difficult to determine the induction period duration since its spectrum extended the region of *700-800 nm* where the signal of explosive luminescence was detected. This result can be referred to CP samples, since the luminescence of SC and ThC samples was less intensive. Similar luminescence signals were found in inert media (white paper, magnesia, opal glass and others). Its nature is not clear, it is to be studied, and spectral structure composition is of primary interest. These kinds of the research are being planned by us.

In radiation by the third harmonic ($\lambda_0 = 354.7$ nm) of silver azide SC at energy density of **H** ≥ *10 mJ/cm²* distinct glow within the time of laser pulse action was observed. This glow could be possibly caused by sample luminescence. At **H** ≥ *35 mJ/cm²* darkening of the radiation zone was observed. Its depth strongly differed for different types of samples and depended on their micro- and macrodefect structure (microcracks, joints and so forth). It was noticed that the more was the depth of the area of darkening, the smaller was IET of the sample. In addition, distinct light scattering occurred in samples with greater depth of darkening. The behavior of silver azide ThC with uncovered surface was generally the same as that for SC. The luminescence signal was reliably detected at **H** ≥ *10 mJ/cm²*. In case of near-surface layer darkening it was recorded at **H** ≥ *30 mJ/cm²*. The distinction was that darkening of the irradiated region occurred in the thin near-surface layer and its thickness could not be measured. It is to be noted that IETs for threadlike crystals (**$H_{05}$** = *600 mJ/cm²*) were considerably higher, than IETs for scaly crystals (**$H_{05}$** = *40 mJ/cm²*). The behavior of silver and lead azide CPs was close to the behaviour of SCs. However, darkening of powders was not revealed due to low IET. Although, IET levels were different for SC, ThC, and CP, basic characteristics of explosive decomposition luminescence (amplitude, luminescence pulse duration and induction period) differed slightly. In conditions of static compressing pressure (covered surface) the distinctions in IETs and characteristics of explosive decomposition became essentially less evident.

In irradiation by the fourth laser harmonic ($\lambda_0 = 266$ nm) initiation with the uncovered surface was high-threshold (**$H_{05}$** = *350-1600 mJ/cm²*) for all the tested types of HMAs. When the explosion of samples failed, there occurred strong darkening of the surface in the region of laser beam action indicating partial or complete decomposition of the sample in the irradiation zone. The surface started to darken at irradiation levels larger than *20 mJ/cm²*. When irradiation levels were lower, there occurred the glow signal characteristic for luminescence. If irradiation levels were higher, the glow signal characteristic for optical breakdown could be observed. It is to be noted that in the latter case a glow flash went with a "clap" and flame-tongue above the sample surface. The threshold of optical breakdown for SC and ThC made about *300 mJ/cm²*, for CP it made approximately 50 *mJ/cm²*. The behavior of all the samples with the covered surface was approximately identical, and initiation can be referred to a low-threshold one.

**Discussion of Results**

To ease the analysis the results of IET measuring for silver azides are shown in diagrams in Fig. 3 (a, b, c).

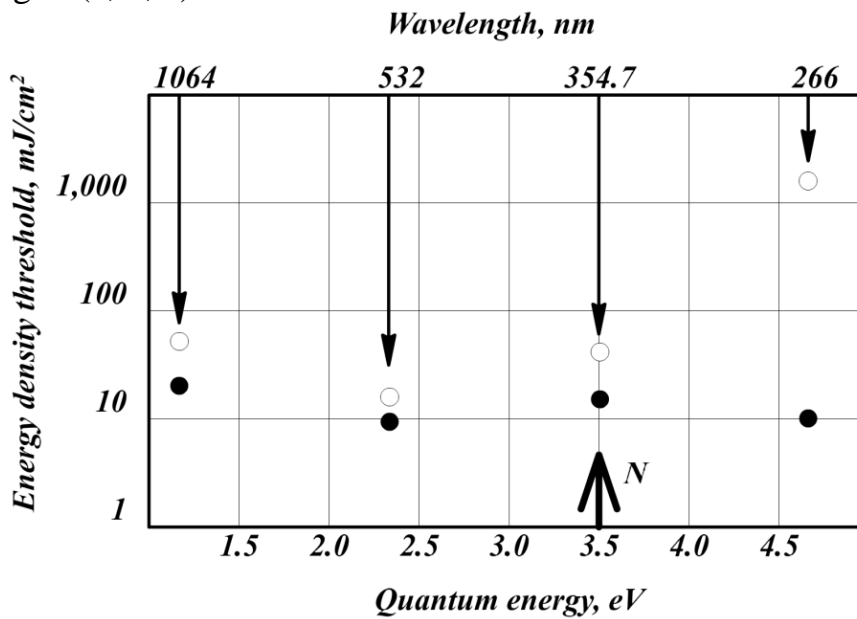

***Fig. 3a.** IET values of AgN$_3$ macrocrystals at different wavelengths of laser radiation (○ is uncovered samples; ● is covered samples)*

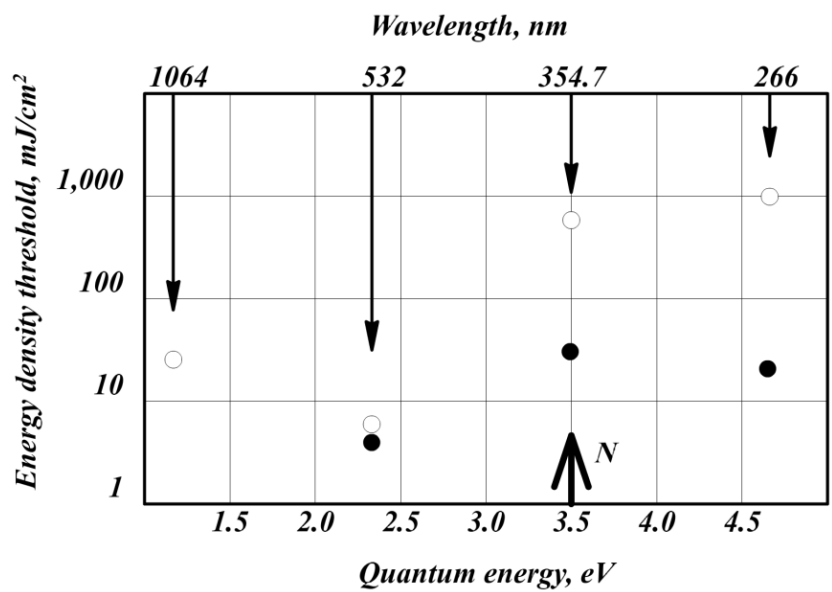

***Fig. 3b.** IET values of threadlike AgN$_3$ crystals at different wavelengths of laser radiation (○ is uncovered samples; ● is covered samples)*

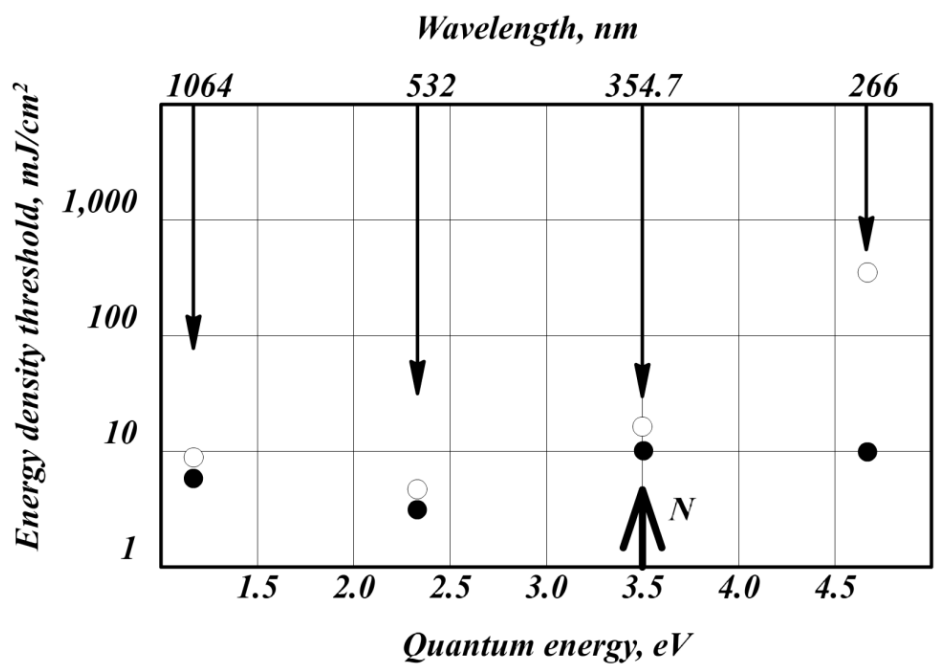

***Fig.3c.*** *IET values of compacted AgN$_3$ powders at different wavelengths of laser radiation*
*(○ is uncovered samples; ● is covered samples)*

In Figures 3 (a, b, c) the photon (quantum) energy is plotted along the abscissa axis, and arrows (on the top of the Figures) indicate the position corresponding to harmonics wavelengths. In the point corresponding to the quantum energy *3.5 eV* there is ***N*** index. Since the bandgap width for AgN$_3$ is considered to be ~*3.5 eV* as well, ***N*** index conditionally separates the transparent region and the region of bandgap absorption.

The results of initiation for Pb(N$_3$)$_2$ are presented in Fig. 4 for comparison.

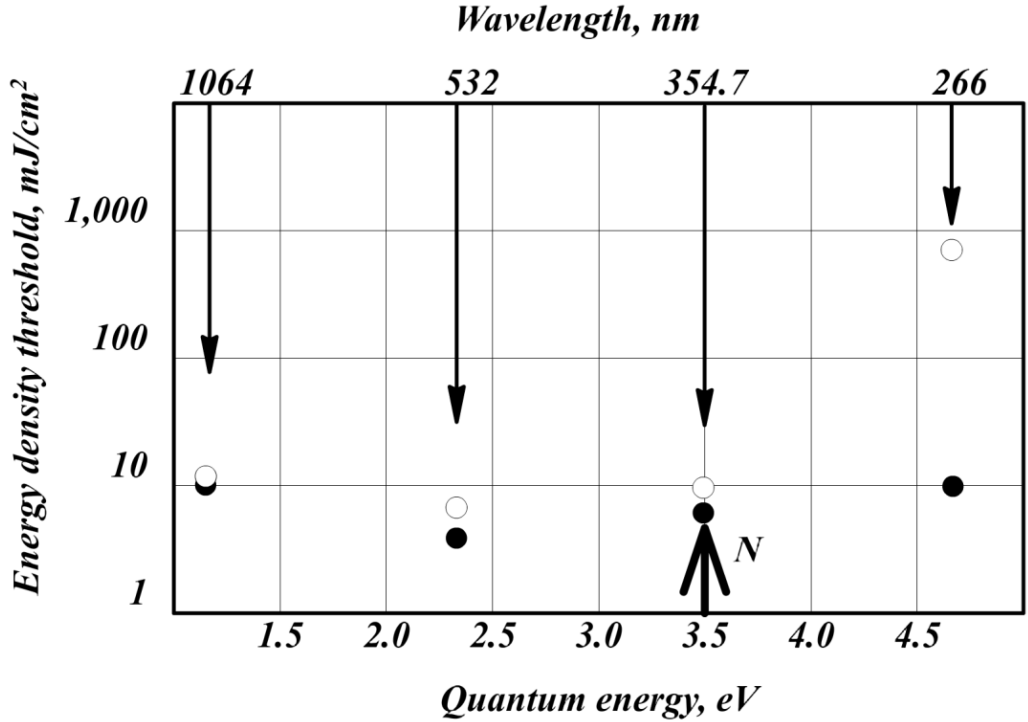

**Fig. 4.** IET values of compacted Pb(N$_3$)$_2$ powders at different wavelengths of laser radiation
(○ is uncovered samples; ● is covered samples)

The interface of the transparent region and the region of bandgap absorption facilitates the analysis of the results. First of all, one can see drop of sensitivity of crystals and the compacted powders in switching from transparent region to the region of bandgap absorption in initiation with the uncovered surface of the sample. The distinctions in IET by the value of $H_{05}$ reach two orders of the value. If recalculate $H_{05}$ values to the averaged absorbed volume energy density $\omega_{05\ av}$ (J/cm$^3$) with $\mu$ values from [6], this distinction will make from 6 to 7 orders for SC and ThC, respectively. This is to be logically explained since these large distinctions in IET cannot be caused by experimental errors. In [6, 7] (initiation of lead azide powders by eximer laser) it is assumed that high thresholds of initiation in the region of bandgap absorption are related to the small depth of radiation penetration in the substance, formation of a high-temperature site in a thin near-surface layer and its fast gas-dynamic unloading through the uncovered surface.

Thresholds measured by us virtually indicate similar development of the process. Thus, at threshold action with the energy density of *1000 mJ/cm$^2$* and absorption index $\mu = 10^5$ cm$^{-1}$ characteristics for HMAs at wavelength of *266 nm* the absorbed volume energy density of LR will make ~$10^5$ J/cm$^3$, that more than one order of the value exceeds the thermal effect of HMA explosive decomposition reaction (~$5 \cdot 10^3$ J/cm$^3$). Overheated plasma layer expands with the velocity which according to measurements [4, 10] makes $v \sim 10^5 - 10^6$ m/s by the moment of the end of the laser pulse, and the time of unloading $\tau_{un} = 1/(\mu \cdot v)$ can make $10^{-10} - 10^{-11}$ s. Thus, despite approximate estimations, it can be confirmed that by the end of LR action ($\tau_p =$

$= 10^{-8}$ s) and during the induction period there can occur intensive gas-dynamic unloading of the site. It is evident that transformation of the thermal site into the explosive decomposition site requires in this situation high levels of heat input from laser acting. It is also evident that the condition of the surface (uncovered-covered) can crucially influence the development of the site. In case with the uncovered surface if wavelength reduces, the *μ* parameter grows, the thickness of the chemically-active layer *h* (*h* ~ *1/μ*) the site develops in reduces, the intensity of gas-dynamic unloading increases through the free surface and IET grows.

When the surface is covered with pressing pressure comparable to the pressure in the site at the stage of its development, gas-dynamic unloading does not practically take place. Therefore IET sharply decreases. These regularities are characteristic for all the materials irrespectively of the technique they are prepared.

In the transparent region of the sample the process develops in an absolutely different way. The behavior of HMAs in initiation with both uncovered and covered surface can be well explained within the limits of the thermal microsite model of ignition. The site develops as a result of cumulation of laser radiation energy in the crystal volume. The cumulation results in the processes stimulating pulses of high level deformations in the sample. These deformations lead to initiation of elementary decomposition acts [11, 12]. The totality of the elementary acts of energy cumulation leads to the nascent of the site. The reaction in the site (on the inner surface of the site region) leads to additional energy release, increase of pressure and temperature inside these sites. The process can increase in an avalanche if the speed of energy release inside the microsite will exceed the speed of energy removal (losses). The characteristic size of this type of microsites as it was shown in [3, 9], makes ~$10^{-5}$ cm. The probability of microsite occurrence is determined by the processes providing cumulation of energy in the process of its release in the sample [1, 11]. On the other hand, the probability of microsite occurrence is determined by some defective (weakened) places in the sample volume which make the centers of energy cumulation and the centers for microsites to occur. The distribution of these centers in the volume of the sample is a statistical property. These centers are typically located deep in the sample volume. Their thermal unloading occurs both into the gas of the developing site and into the medium around the site. This causes increase of the shock elastic matrix deformation and therefore, to the increase of the probability of explosive decomposition. In this case the condition of the surface (uncovered-covered) weakly influences the development of this type of sites, and consequently, it weakly affects the explosive process. Therefore, in excitation by radiation pulses of the first and second harmonics low-threshold initiation takes place all the time.

Characteristics of initiation by the third laser harmonic of the neodymium laser are considered to be of interest. In excitation with the uncovered surface a wide spread of IET values for samples of different origin can be observed. In this case the radiation wavelength is close to the edge of the bandgap width. The type of the absorption spectrum in this region is determined by material imperfection and its structure. The researches in the wavelength region of the bandgap absorption edge are known to be used for identification of optical materials imperfection. We assume that the spread of IET values for samples of different origin is related to the distinction of radiation absorption of the third harmonic in these samples. The following fact observed in determination of IET can confirm this conclusion. In case when initiation in excitation of the sample does not occur, in the region of action on the sample there can be observed characteristic darkening caused by partial HMA decomposition that does not run to explosion. It was noticed that the more the depth of darkening of the sample, the smaller IET of the sample was. Therefore, there was a wide spread of IET values (large probabilistic explosion interval).

**Conclusions**

- It has been experimentally found that in switching from the transparent region (the first and the second harmonics of YAG:$Nd^{3+}$-laser) in the region of bandgap absorption (the fourth laser harmonic) the initiation thresholds for HMAs with the uncovered surface of samples jump. It has been shown that strong distinction in character of unloading from the excitation zone (from heat conductive unloading in the first case to a gas-dynamic one in the second);
- It has been found that in initiation by the third laser harmonic the behavior of HMAs is ambiguous: both low-threshold and high-threshold initiation can be observed. The assumed cause of this behavior is that the radiation wavelength of the third laser harmonic is located near to the edge of HMA bandgap absorption, which position can shift for different samples. Darkening of the samples caused by partial decomposition in irradiation by the third laser harmonics with the energies lower than the threshold ones found in the experiment proves the assumption. The degree and value of sample darkening are found to be different for different samples;
- Covering the irradiated surface with a transparent dielectric with the compressive force commensurable with the pressure of explosive decomposition leads to a sharp decrease of gas-dynamic unloading and, as consequence, to realization of a low-threshold mode of initiation in all the studied interval of wavelengths;
- The results of the research can be interpreted within the limits of microsite model of ES ignition.

# References

1. Brish A.A., Galeev I.A., Zaitsev B.N., Sbitnev E.A., Tatarintsev L.V. Mechanism of initiation of condensed explosives by laser radiation // Combustion, Explosion, and Shock Waves, Vol. 5, No. 4, pp. 326-328, 1969
2. Yang L.C., Menichelli V.J. Detonation of insensitive high explosives by a Q-switched ruby laser // Appl. Phys. Lett., Vol. 19, No. 11, 1971
3. Aleksandrov E.I., Voznyuk A.G. Initiation of lead azide by laser radiation // Combustion, Explosion, and Shock Waves, V. 14, No. 4, pp. 480-483, 1978
4. Aleksandrov E.I., Tsipilev V.P. Size effect in the initiation of pressed lead azide by single-pulse laser radiation // Combustion, Explosion, and Shock Waves. Vol. 17, No. 5, pp. 550-552, 1981
5. Korepanov V.I., Lisitsyn V.M., Oleshko V.I., Tsipilev V.P. To the Question about Kinetics and Mechanism of Explosive Decomposition of Heavy Metal Azides // Combustion, Explosion, and Shock Waves, Vol. 42, No. 1, pp. 94-106, 2006
6. Aleksandrov E.I. HMA Initiation Research under UV-laser Radiation // Kvant. Electr. Pril, Vol. 8, pp.32-36, 1977
7. V.M. Lisitsyn, V.I. Oleshko, V.P. Tsipilev, A.N. Yakovlev. About Power Thresholds, Criteria, Kinetics and Mechanisms of Ignition of Explosives by Laser Pulses and Pulsed Electron Beams // Russian Physics Journal, No. 10, pp. 200-203, 2006
8. Alexandrov E.I., Zykov I.Yu., Morozova E.Yu., Oleshko V.I., Losev V.F., Panchenko Yu.N., Tsipilev V.P., Yakovlev A.N. Research of Explosion Decomposition of Heavy Metal Azides under Excimer Laser Radiation and Radiation of Carbon Dioxide Laser // Fundamentalnye i prikladnye problema sovremennoy mekhaniki (Collected Papers of Conference), Tomsk, Publ. of TSU, 2008, pp. 37-38
9. Kriger V.G., Kalensky A.V., Kolbasov S.V., Konkov V.V., Plusnin V.F. Pulsed Luminescence of Silver Azide Initiated by Excimer Laser // Thesises of 9th International Conference "Radiation Physics and Chemistry of Inorganic Materials", Tomsk, Publ. of TSU, 1996, pp. 222-223
10. Aleksandrov E.I., Tsipilev V.P. Influence of the mode structure of laser radiation on lead azide stability // Combustion, Explosion, and Shock Waves, Vol. 19, No. 4, pp. 505-508, 1983
11. Tsipilev V.P., Lisitsyn V.M., *Damamme G., *Malys D. BLAST INITIATION OF HEAVY METAL AZIDES BY LASER PULSE RADIATION IN UV SPECTRAL FIELD // Russian Physics Journal, No. 8/2, pp.320-323, 2009
12. Lisitsyn V.M., Zhuravlev Yu.N., Oleshko V.I., Fedorov D.G., Tsipilev V.P. Deformation mechanism of the explosive degradation of heavy metal azides under pulsed treatments // High Energy Chemistry, Vol. 40, No. 4, pp. 218-223, 2006
13. Lisitsyn V.M., Zhuravlev Yu.N., Oleshko V.I., Fedorov D.G., Tsipilev V.P. // Chem. Phys. Journal, Vol. 25, No. 2, pp. 59-64, 2006